\documentclass{aastex701}

\usepackage{graphicx}
\usepackage{lipsum}
\usepackage{caption}
\usepackage{url}
\usepackage{hyperref}
\usepackage{amsmath}
\usepackage{float}
\numberwithin{equation}{section}
\usepackage{amssymb}
\usepackage{rotating}
\usepackage[export]{adjustbox}
\usepackage{xcolor}
\usepackage{booktabs}

\hypersetup{
    colorlinks=true,
    linkcolor=blue,
    citecolor=blue,
    urlcolor=blue
}

\begin{document}

\title{Collective Vortex Dynamics: From Isolated Vortices to their Communities}

\author[0009-0000-4129-1324]{Lauren McClure}
\affiliation{Plasma Dynamics Group, 
School of Electrical and Electronic Engineering, 
The University of Sheffield, Sir Frederick Mappin Building, Mappin Street, Sheffield, S1 3JD}
\email{lmcclure1@sheffield.ac.uk}

\author[0000-0001-5414-0197]{Suzana Silva}
\affiliation{Plasma Dynamics Group, 
School of Electrical and Electronic Engineering, 
The University of Sheffield, Sir Frederick Mappin Building, Mappin Street, Sheffield, S1 3JD}
\email{suzana.silva@sheffield.ac.uk}

\author[0000-0002-9546-2368]{Gary Verth}
\affiliation{Plasma Dynamics Group, 
School of Mathematical and Physical Sciences, 
The University of Sheffield, Hicks Building, Hounsfield Road, Sheffield, S3 7RH}
\email{g.verth@sheffield.ac.uk}

\author[0000-0002-3066-7653]{Istvan Ballai}
\affiliation{Plasma Dynamics Group, 
School of Mathematical and Physical Sciences, 
The University of Sheffield, Hicks Building, Hounsfield Road, Sheffield, S3 7RH}
\email{i.ballai@sheffield.ac.uk}
 
\author[0000-0002-0893-7346]{Viktor Fedun}
\affiliation{Plasma Dynamics Group, 
School of Electrical and Electronic Engineering, 
The University of Sheffield, Sir Frederick Mappin Building, Mappin Street, Sheffield, S1 3JD}
\email{v.fedun@sheffield.ac.uk}

\begin{abstract}

Small-scale vortical motions in the upper solar atmosphere are abundant and occupy about 2.8\% of the photosphere at any given time. Although considerable work has focused on the detection and analysis of individual solar vortices, the interconnected and multi-scale behaviour of these coherent structures remains largely unexplored. We present a methodology for studying this behaviour through vortex interactions, to improve our understanding of how small- and large-scale photospheric flows contribute to energy transfer into the upper solar atmosphere and to the driving of solar activity. We represent vortices as a network of interacting structures. We apply a community detection algorithm to derive an optimal reduced network composed of highly interconnected vortex groups. From the interaction patterns and group structure, we define three roles within each community: peripheral, connector and hub. We then track both vortex communities and their member vortices from the photosphere into the chromosphere and across their lifetimes. On average, vortices assigned to these roles persist to greater heights in the chromosphere and have longer lifetimes than unclassified vortices. This shows that community detection can identify vortices with greater dynamical influence on the upper solar atmosphere. We also find that 32\% to 58.6\% of vortex communities exhibit global periodic behaviour following a helical path. This collective vortical motion may indicate an enhanced mechanism for wave excitation. Solar vortical community detection, therefore, offers a new framework for studying solar vortices and a new perspective on the importance of collective vortex dynamics.

\end{abstract}

\section{Introduction} \label{sec:intro}
Vortical motions are observed throughout the solar atmosphere, from photospheric swirls to large-scale chromospheric tornados \citep{Wedemeyer_2012, Tziotziou_2023}, and have been proposed to transport energy \citep{Yadav_2020, Kuniyoshi_2023, Silva_2024a, Silva_2024b}, mass, and momentum into the upper atmosphere, with implications for coronal heating and solar wind acceleration \citep{Zuccarello_2017, Finley_2022, Tziotziou_2023}. Characterising these dynamics is therefore important for understanding upper atmospheric energetics, though quantifying small-scale vortex effects remains challenging given their abundance and complex dynamics \citep{Giagkiozis_2018}.
Much of the solar vortex literature focuses on determining vortex sizes, lifetimes, occurrence rates, and associated energy and transport estimates \citep{Moll_2011, Shelyag_2011b, Vargas_2011, Giagkiozis_2018, Tziotzou_2018, Aljohani_2022, McClure_2026}, with many studies also examining connections to wave excitation, magnetic concentrations, and localised heating \citep{Wedemeyer_2012, Tziotziou_2023, Kuniyoshi_2025}. However, these statistical approaches predominantly characterise vortices as individual features rather than as collective populations.
While this perspective can be appropriate for isolated structures, this is not the case for photospheric vortices, considering they appear largely in groups \cite{Giagkiozis_2018, Silva_2020} and can therefore interact collectively, generating larger-scale dynamics. To address this aspect, we introduce a community detection framework that identifies coherent, interacting vortex groups from an interaction network. This enables community-level observables and a new view of collective dynamics.



Network science reduces complex systems into group dynamics \citep{Newman_2010}, with modularity maximisation partitioning networks based on node interactions \citep{Newman_2004, Newman_2006}. This approach has been widely applied across neuroscience \citep{Bassett_2011_Neuro}, economics \citep{MacMahon_2015_Finance}, and biological networks \citep{Lewis_2010_Biology}. In hydrodynamics, network and graph-theoretic methods have been applied to turbulent vortex systems and unsteady aerodynamic modelling \citep{Nair_Taira_2015, Taira_2016, Gopalakrishnan_Meena_2018}. \citet{Nair_Taira_2015} introduced a vortex network representation where interaction strength is defined via induced velocity, consistent with the Biot–Savart formulation \citep{Saffman_1992}.
Treating vortices as interacting structures rather than isolated features has advanced our understanding of vortex-driven systems. One of the first explicit community detection frameworks for vortical flows was proposed by \citet{Gopalakrishnan_Meena_2018}. They showed that macroscopic hydrodynamic dynamics can be captured using a reduced description based on vortex communities. Their approach enables the prediction of selected observables and is reported to be robust to noise and turbulent variability. Building on this, \citet{Gopalakrishnan_Meena_2021} classified vortices and communities according to their intra-community and inter-community interactions. They found that the network-based categories align with distinct physical flow characteristics.

Vortical community detection remains a relatively recent development, yet it has already provided insights into the global dynamics of turbulent systems and demonstrates clear potential for further progress in hydrodynamics \citep{Taira_2016, Gopalakrishnan_Meena_2018}. These methods are robust in the presence of noise and turbulence, while scaling effectively to large and complex systems \citep{Gopalakrishnan_Meena_2018}. Collectively, these properties motivate the application of vortex community detection to the solar atmosphere, where vortices are abundant, and their mutual interactions may play an important role in determining their overall dynamical and energetical impact. Extending these techniques to magnetohydrodynamic environments introduces additional challenges, including compressibility effects, magnetic forces and the need to account for the full three-dimensional structure of solar vortical flows \citep{Priest_2014}. 


We introduce a solar vortical community detection framework, applying it to simulated quiet-Sun data spanning the photosphere to chromosphere \citep{Gudiksen_2011}. By constructing a vortex interaction network, we identify dynamically influential vortices and link them to distinct physical properties. We then apply modularity maximisation to detect coherent vortex communities, revealing how their structure relates to collective motion evolving across time and atmospheric height

\section{Methodology} \label{sec:Gamma}
\subsection{Bifrost Simulation Data} \label{sec:Bifrost}
The dataset used in this study was generated with the multidimensional radiative magnetohydrodynamics (MHD) code Bifrost \citep{Gudiksen_2011}, which simulates realistic photospheric velocity and magnetic fields. In this paper, we use the simulation `ch024031\_by200bz005', which simulates coronal hole conditions and is publicly available through the Hinode Science Data Center. Further details of the simulation setup are given by \citet{Gudiksen_2011}, \citet{Carlsson_2016} and \citet{Finley_2022}.

We analyse the full horizontal extent of the simulation ($24 \times 24$~Mm) at a horizontal resolution of 31~km \citep{Carlsson_2016, Finley_2022}. The analysis uses a continuous $\approx 37$ minute time series (indices 249 - 469) sampled every 10s. The vertical grid is non-uniform, and we consider heights from $\approx 0.45$~Mm above the solar surface (photospheric layers) up to $\approx 2$~Mm to assess vortex-tube heights and coherent structures extending into the upper atmosphere. We choose the initial height of $\approx 0.45$~Mm because vortical structures are larger than in the lower photosphere and thus better resolved by the horizontal grid, improving the robustness of community detection.

Across this interval, the vertical spacing varies between 12.35 and 12.7~km. The high spatial resolution and cadence make the dataset well-suited for studying and tracking small-scale turbulent motions such as solar vortices.

\begin{figure*}[t!]
  \centering
  \captionsetup{font=small,skip=2pt}
  \begin{adjustbox}{max width=\textwidth, max totalheight=1\textheight}
    \includegraphics[width=10\linewidth]{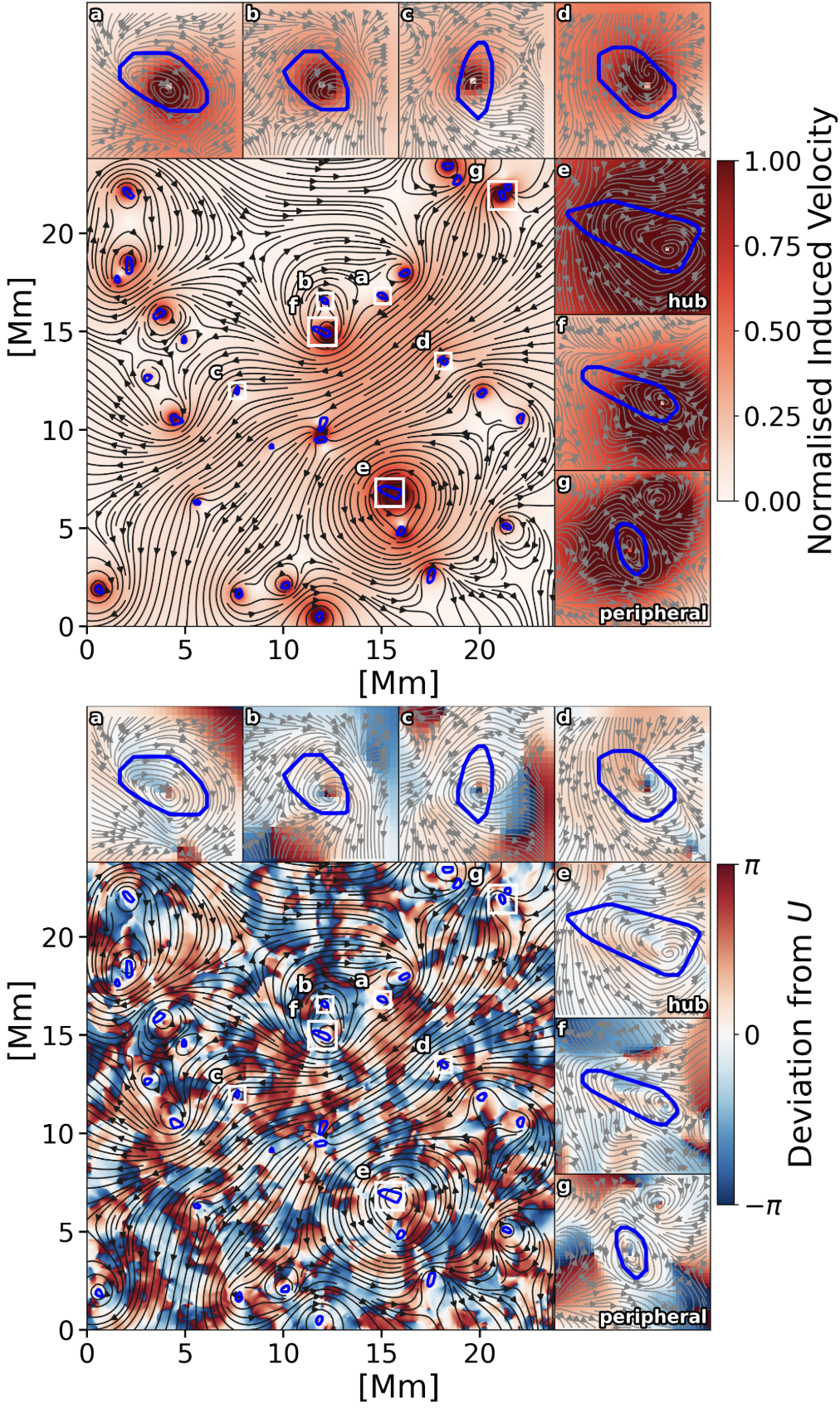}
  \end{adjustbox}
  \caption{Vortex detection results over the full field of view at $t=0$ seconds. Black streamlines in the main panels show the total induced velocity field, computed from the contributions of all detected vortices, while panels a–g show streamlines in grey of the actual simulated velocity fields. Panels e and g highlight the hub and peripheral vortices at this time frame. The top panel shows the normalised induced velocity magnitude, and the bottom panel shows the angle (in radians) between the velocity field and induced velocity vectors.}
\label{fig:Ind_Vel}
\end{figure*}

\subsection{Vortex Interactions}\label{sec:vortex_interactions}

In this work, we apply a community detection algorithm to systems of solar vortices \citep{Newman_2006,Blondel_2008}. Similar approaches framework have been employed to hydrodynamic point-vortex systems (idealised, two-dimensional flow model in which all vorticity is concentrated at a single point with finite circulation) by \citet{Nair_Taira_2015}, \citet{Taira_2016} and \citet{Gopalakrishnan_Meena_2018}, who demonstrated that community-based modelling can capture aspects of global flow evolution and provide a route toward predicting collective motion and observable signatures. Here, we extend this concept to a magnetohydrodynamic setting by detecting vortices in the Bifrost simulation described in Section \ref{sec:Bifrost} using the $\Gamma$ method \citep{Graftieaux_2001}. We adopt a dual $\Gamma$ approach with dynamic thresholding to reduce false positives and to track vortices more robustly across their full spatial and temporal extent. Similar to the study by \citet{McClure_2026}, we apply a normalisation pre-processing step. All vortex detection details used in this work are reported in Section \ref{sec:vortex_detection}.

We quantify the rotational strength enclosed by the detected boundary of vortex $i$ through its circulation,
\begin{equation}
\gamma_i = \sum_{V_i} \omega_z \Delta A,
\end{equation}
where $V_i$ denotes the region enclosed by the boundary of the $i$-th vortex, $\omega_z$ is the vertical component of vorticity and $\Delta A$ is the area of a grid cell.

The circulation, $\gamma_i$, is then used to define the velocity induced by vortex $i$ to vortex $j$ ($i \neq j$) as
\begin{equation}
u_{i \rightarrow j} = \frac{\gamma_i}{2\pi |\boldsymbol{r}_j - \boldsymbol{r}_i|},
\end{equation}
where $\boldsymbol{r}_i$ and $\boldsymbol{r}_j$ denote the horizontal positions of vortex centres \citep{Saffman_1992, Nair_Taira_2015, Gopalakrishnan_Meena_2018}. For $i=j$, we set $u_{i \rightarrow j}=0$ in all subsequent calculations, as self-induction is undefined within this two-dimensional vortex interaction formulation. In this approach, $\gamma_i$ scales with the horizontal area enclosed by the detected vortex boundary, implying that the induced velocity is sensitive to the chosen boundary definition. 

The velocity induced by one vortex on another depends on their relative positions. By translating the coordinate system so that a source vortex is located at the origin, its induced velocity can be expressed at any point in the surrounding flow. In this induced field, streamlines form concentric circles centred on the vortex. The flow speed is uniform along each circle but varies between circles, and the motion is purely tangential, with no radial component \citep{Saffman_1992}. Applying this construction to all vortices in the domain yields the total induced velocity field, which represents the collective influence of the vortex population on the surrounding flow. A visualisation of this field is shown in Fig.~\ref{fig:Ind_Vel}. 
The black streamlines illustrate the total induced velocity field, highlighting the net dynamical influence of individual vortices. We also present the induced velocity magnitude and its angular deviation from the actual velocity vector. Regions surrounding vortex centres show strong directional agreement between these fields, and  vortex's region also coincides with the locations of the largest induced velocity magnitude. Note that the induced velocity derives from detected vortices only, so undetected vortices may introduce discrepancies reflecting detection limitations. Given the requirements on vortex strength by the gamma method, any missed vortices are expected to be smaller and rotationally weaker than those detected, meaning their circulation and contribution to the total induced velocity field would likely be small.

\begin{figure*}[t!]

  \centering
  \captionsetup{font=small,skip=2pt}
  \begin{adjustbox}{max width=\textwidth, max totalheight=1\textheight}
    \includegraphics[width=10\linewidth]{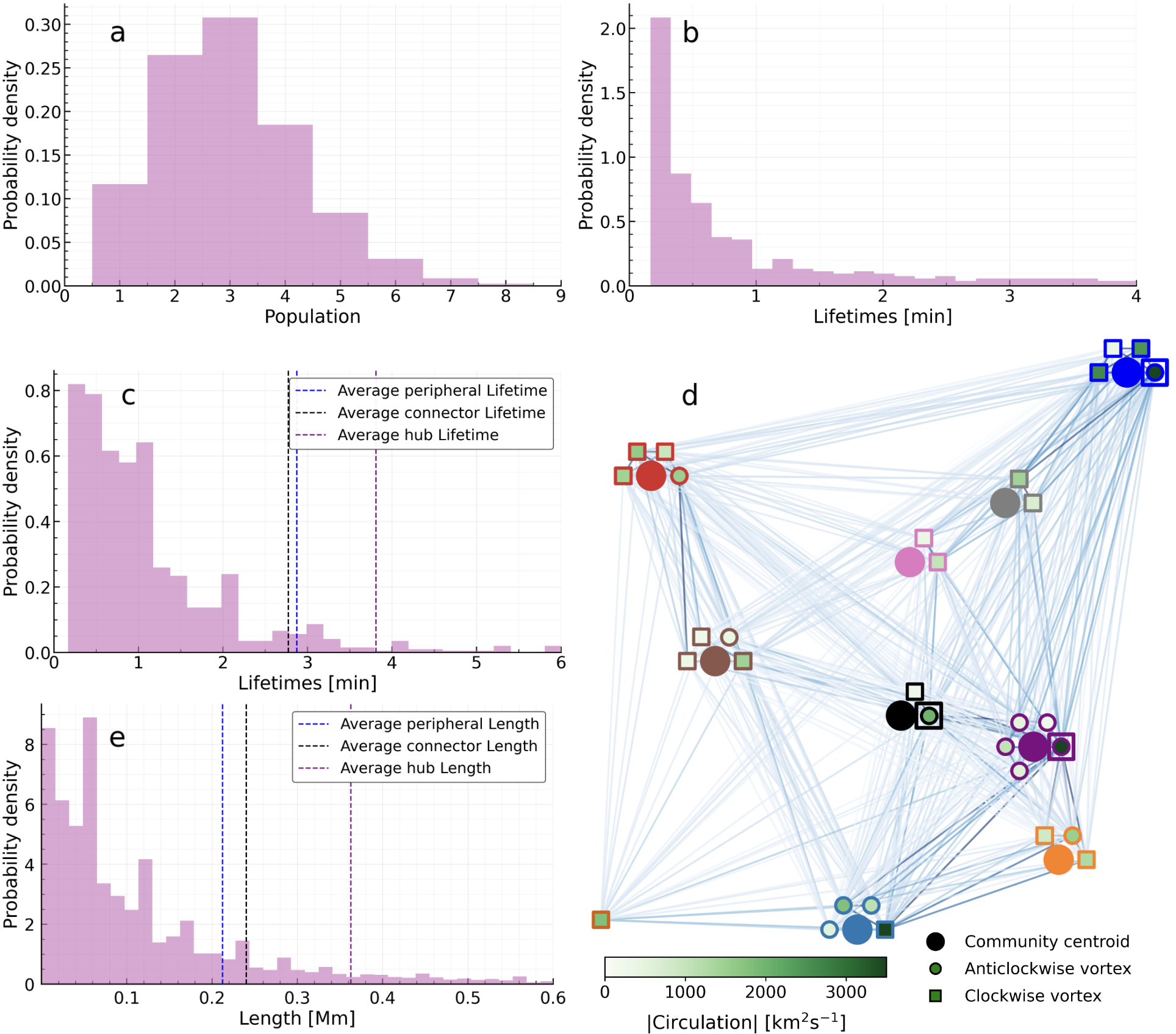}
  \end{adjustbox}
  \caption{
  Histograms of property distributions for vortex communities (panels a, b) and individual vortices (panels c, e). Panels a and b show community size and duration. Panel e shows vortex tube length, and panel c shows vortex lifetime. Panel d shows an example community structure, where the centroid colours and vortex border colour indicates community membership, with each centroid surrounded by its member vortices. Marker faces are coloured by $|\gamma_i|$ and border colour indicates community membership. Vortices are linked by lines coloured by the mean induced velocity magnitude between each pair. Boxes highlight the peripheral, connector and hub (blue, black and purple).
  }
\label{fig:Histograms}
\end{figure*}

The induced velocity interactions, however, provide a natural framework for representing the vortex population as a dynamically connected system, rather than a collection of isolated features. We construct an interaction network based on these induced velocities and apply modularity maximisation to identify community structures within the vortex population. We present an example of a community structure for our data in Fig.~\ref{fig:Histograms} (d) \citep{Newman_2006,Gopalakrishnan_Meena_2018}. The network partition is obtained using an iterative Louvain method applied to the induced velocity connectivity \citep{Blondel_2008}. This approach reveals groups of vortices that are preferentially interconnected and enables a classification of vortices according to their structural role within the community structure. 

Using this framework, we define three classes of influential vortices in the network: hub, peripheral and connector vortices \citep{Gopalakrishnan_Meena_2021}. In the induced velocity network, a hub vortex contributes the largest share of the induced velocity and is typically the most influential vortex in the system. Figure \ref{fig:Ind_Vel} illustrates this behaviour for the hub vortex (e), which produces the strongest local distortion of the surrounding streamlines.

Peripheral vortices, in contrast, generate the largest induced velocity within their own community but belong to groups that have relatively weak inter-community interaction in the reduced network. As a result, they exert a strong local influence while contributing less to the global network dynamics. In Fig.~\ref{fig:Ind_Vel}, the peripheral vortex (g) is located within the cluster in the upper-right region, where the associated streamlines indicate that the induced flow remains largely confined to the local group. 

Connector vortices, in contrast, are associated with strong interactions that extend across multiple communities \citep{Gopalakrishnan_Meena_2021}. They are characterized by high intra-community connectivity within their own group, while simultaneously maintaining comparatively strong inter-community links. As a result, connector vortices play an important role in mediating interactions between otherwise weakly coupled communities. 

\section{Results}
\subsection{Lifetimes of Photospheric Vortices}

Using the dual part $\Gamma$ method described in Section \ref{sec:vortex_detection}, we analysed the full domain and identified more than 1000 coherent vortices at a fixed photospheric height of $\approx 0.45$~Mm. Vortices were tracked across 220 frames, and lifetimes were recorded for those that both formed and fully dissipated within the analysed time series. Each vortex contributes a single entry to the statistics of its assigned class, even if the class classification persists across multiple frames. As shown in Fig.~\ref{fig:Histograms} (c), the three vortex classes show longer mean detected lifetimes than the full population: 2.87 minutes for peripheral vortices, 2.77 minutes for connector vortices and 3.81 minutes for hub vortices, compared to 1.18 minutes for the full sample. Mean values are computed only from fully dissipated vortices.

\begin{figure*}[t!]

  \centering
  \captionsetup{font=small,skip=2pt}
  \begin{adjustbox}{max width=\textwidth, max totalheight=1\textheight}
    \includegraphics[width=10\linewidth]{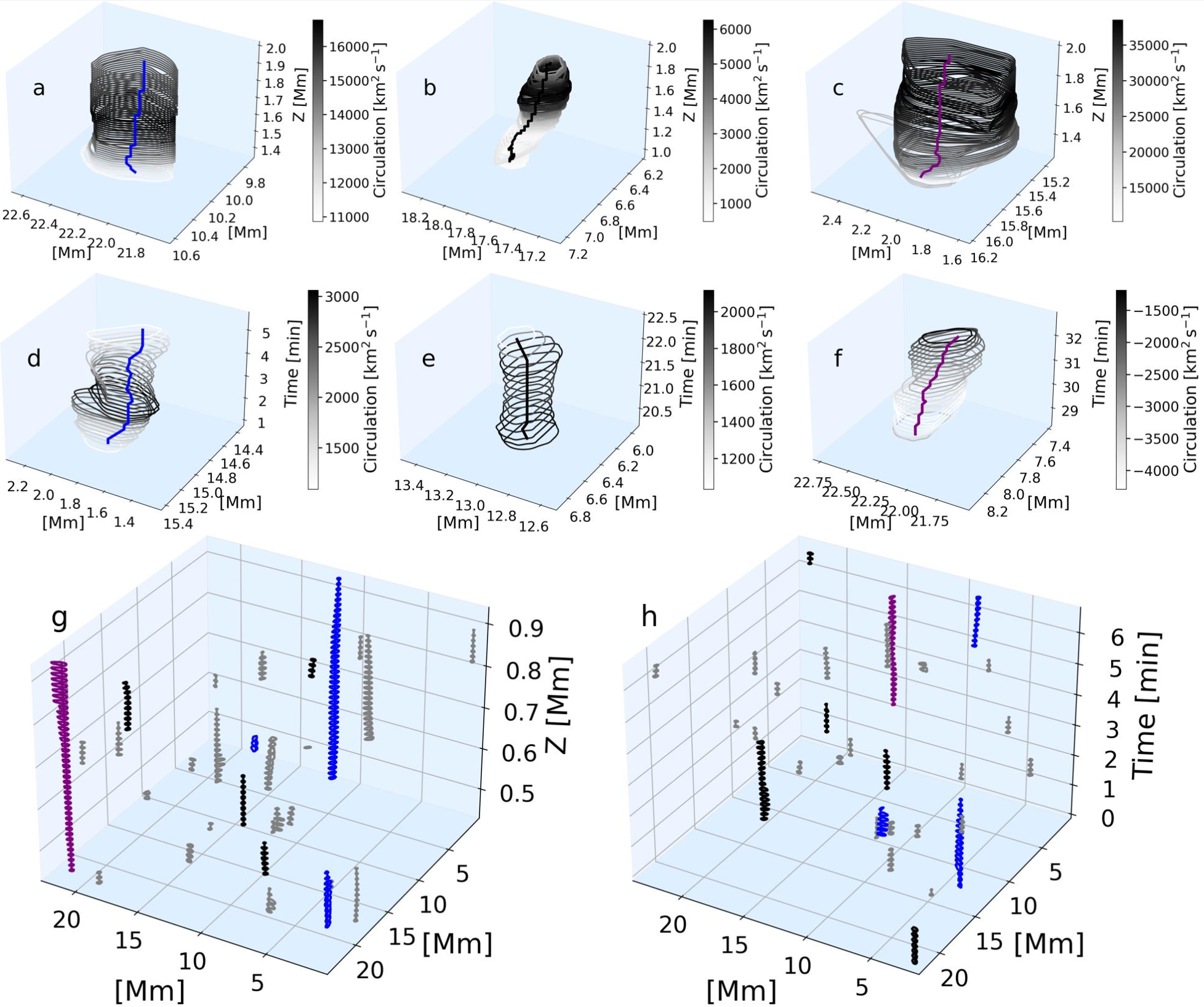}
  \end{adjustbox}
  \caption{Examples of vortex detections over their vertical extent or lifetime. Figures a-f show individual vortex detections, where the boundaries throughout their height range or lifetime are coloured by the circulation at that height/time. For these panels, the detected centre is coloured by vortex role - blue for peripheral, black for connector and purple for hub. This colour scheme is also used in figures g and h to represent the classified vortices, in which grey is any vortex that does not fall into any of the three classes. Panels g and h show a subset of detected vortices at different heights or times.}
\label{fig:3D_TH}
\end{figure*}

Figure \ref{fig:3D_TH} presents examples of vortex evolution over time (d, e, f and h). The individual vortex tracks shown in panels d, e and f correspond to representative peripheral, connector and hub vortices, respectively. Circulation is displayed at each time step to illustrate variations in rotational strength during the vortex evolution and decay. The peripheral vortex in panel d begins with relatively weak circulation, strengthens steadily during its development, and then weakens as it dissipates. The connector vortex shows a more pronounced change in net rotational strength mainly during its decay phase, coinciding with a reduction in vortex area. The clockwise-oriented hub vortex shows a gradual decrease in circulation magnitude over its approximately four-minute detected lifetime, while its boundary undergoes noticeable fluctuations.

Panel h shows a subset of vortices detected over an $\approx 7$~minute interval, with the peripheral, connector and hub vortices highlighted in blue, black and purple, respectively. The presented examples suggest that unclassified vortices are typically shorter-lived, whereas vortices belonging to one of the three classes tend to persist for longer durations, consistent with the statistical trends shown in Fig.~\ref{fig:Histograms}.

Longer-lived vortices may exert a greater cumulative influence on the surrounding plasma simply because they remain coherent for longer. Photospheric vortices have been proposed as seeds for vortex tubes that extend into higher atmospheric layers and may facilitate energy into the chromosphere and corona \citep{Shelyag_2011b,Wedemeyer_2012,Breu_2023}. In this context, peripheral, connector and hub vortices, which remain coherent for longer than the typical vortex in our sample, could contribute more significantly to upward energy transport. However, quantifying this contribution requires a direct energy flux analysis. 

Role classes likely also describe how a vortex is coupled to its environment, and that coupling can directly affect how long a vortex remains detectable and coherent. This coupling can help determine the balance between processes that generate or maintain a vortex and processes that promote decay \citep{Tziotziou_2023}. Each role has different characteristics that are defined in terms of vortex interaction; however, these differences often also have implications for where the vortex sits relative to other vortices, and therefore for which processes are likely to dominate the local flow \citep{Gopalakrishnan_Meena_2021}.  


Peripheral vortices tend to occupy spatially isolated communities, shielding them from disruptive interactions and strain from neighbouring groups. Connectors are more spatially central, exposing them to ongoing interactions that may help regenerate and sustain them. Hub vortices, defined by the largest total induced velocity output, likely combine strong circulation with a central position, benefiting from similar ongoing driving. A hub's larger size and stronger circulation make it harder to disrupt, potentially contributing to a longer detectable lifetime. Connector vortices may persist for analogous reasons. We note that our results reflect continued detection, so vortices maintaining a sufficiently strong circulatory strength may do so partly because their network role and spatial positioning provide a more stable dynamical environment.


\subsection{Photospheric-Chromospheric Vortex Tubes}

Following the vortex lifetime analysis, we also examine the vertical persistence of vortices, measuring how far they extend from the photosphere into the chromosphere before dissipating. We define dissipation as the absence of a local detection in the next horizontal layer. The analysis uses the full horizontal field of view and spans a vertical extent of $\approx 1.55$~Mm. We track vortex tubes rooted at the top of the photosphere ($\approx 0.45$~Mm) over $\approx 37$ minutes of solar time. Across this interval, we detect and track more than 5100 vortex tubes, from tubes confined to the photosphere to those extending into the chromosphere. The mean tube length is 0.16~Mm.

Figure \ref{fig:Histograms} (e) shows the distribution of vortex tube lengths. The full distribution is left-skewed, indicating that most tubes are short. When we restrict the sample to hub vortices, the distribution becomes broader and closer to uniform, consistent with their larger mean tube length of 0.36~Mm. Peripheral and connector vortices also have longer mean lengths of 0.21~Mm and 0.24~Mm, respectively, compared with 0.16~Mm for unclassified vortices.

Each vortex tube contributes one tube length measurement to the statistics of its assigned class, even if it is detected across multiple discrete $z$ heights. We assign the peripheral, connector or hub label at the photospheric base layer and retain this label when tracking the tube’s vertical extent. The class labels shown in Fig.~\ref{fig:Histograms} and the mean values reported here therefore refer to the base layer classification.

Vortex tubes that remain detected at the highest height considered ($\approx 2$~Mm) are excluded from Fig.~\ref{fig:Histograms} (e) and from the mean values in this section, since their full lengths are not observed within the analysed range. Out of the 255 vortices that persist to the final vertical extent, the number of which are part of the highlighted groups are: 12 peripheral, 21 connector and 54 hub vortices, each exceeding 1.5~Mm in length. We exclude these cases to avoid skewing the distribution and to ensure the reported statistics fully reflect observed tube lengths.

Tiles a, b, c and g in Fig.~\ref{fig:3D_TH} show examples of $\Gamma$ detected vortex tubes. We plot the vortex boundaries and centres at each height where the tube is detected. The boundary colour indicates circulation at each height and therefore shows how the net rotational strength varies along the vertical extent. Vortex a is a peripheral vortex that strengthens near the base and maintains relatively high circulation up to its final detected height of $\approx 2$~Mm. The hub vortex (c) shows a similar evolution, but with a shorter detected height of around 0.6~Mm. Of the three examples, the connector vortex (b) is the longest, reaching $\approx 1$~Mm. It has lower circulation in the lower half of the tube, before strengthening as the boundary expands. Towards the upper end of vortex b, both the boundary size and circulation decrease until the vortex is no longer detected. Across the three classes, hub vortices generally show the largest net rotational strength.

Panel g in Fig.~\ref{fig:3D_TH} shows a subset of detected vortex tubes, with the peripheral, connector and hub examples highlighted in blue, black and purple. In this subset, the two tubes with the greatest vertical extents are classified as peripheral or hub. While Fig.~\ref{fig:Histograms} and Fig.~\ref{fig:3D_TH} differ in what they summarise, both evidence that vortices assigned to one of the three classes are more likely to persist to greater heights than unclassified vortices, with the effect most pronounced for hub vortices.

\subsection{Global Trajectories of Vortex Communities }

We identified and tracked 340 distinct vortex communities over the whole time period at $\approx$ 0.45~Mm above the solar surface. Each community represent a different composition of vortices, they contain different numbers of vortices, and they may be instantaneous or span multiple time frames. We report the population (a) and duration of detection (b) for the communities in Fig.~\ref{fig:Histograms}, these distributions are informed by the induced velocity vortex interactions and represent the optimum reduced network structure.

Vortex communities evolve, and their membership changes, vortices move, and circulation varies, all influencing the inferred community position. To quantify community location and motion, we assign each community a representative centre. The centroid of a community is defined as the circulation weighted spatial average of its member vortices \citep{Gopalakrishnan_Meena_2018, Gopalakrishnan_Meena_2021}:
\begin{equation}
\boldsymbol{c}_i = \frac{\sum_{j \in C_i} |\gamma_j|\boldsymbol{r}_j}{\sum_{j \in C_i} |\gamma_j|}.
\end{equation}
The weight is the absolute circulation $|\gamma_j|$. Here, $C_i$ denotes community $i$ and $\boldsymbol{r}_j$ is the position of vortex $j$.

\begin{figure*}[b!]

  \centering
  \captionsetup{font=small,skip=2pt}
  \begin{adjustbox}{max width=\textwidth, max totalheight=1\textheight}
    \includegraphics[width=10\linewidth]{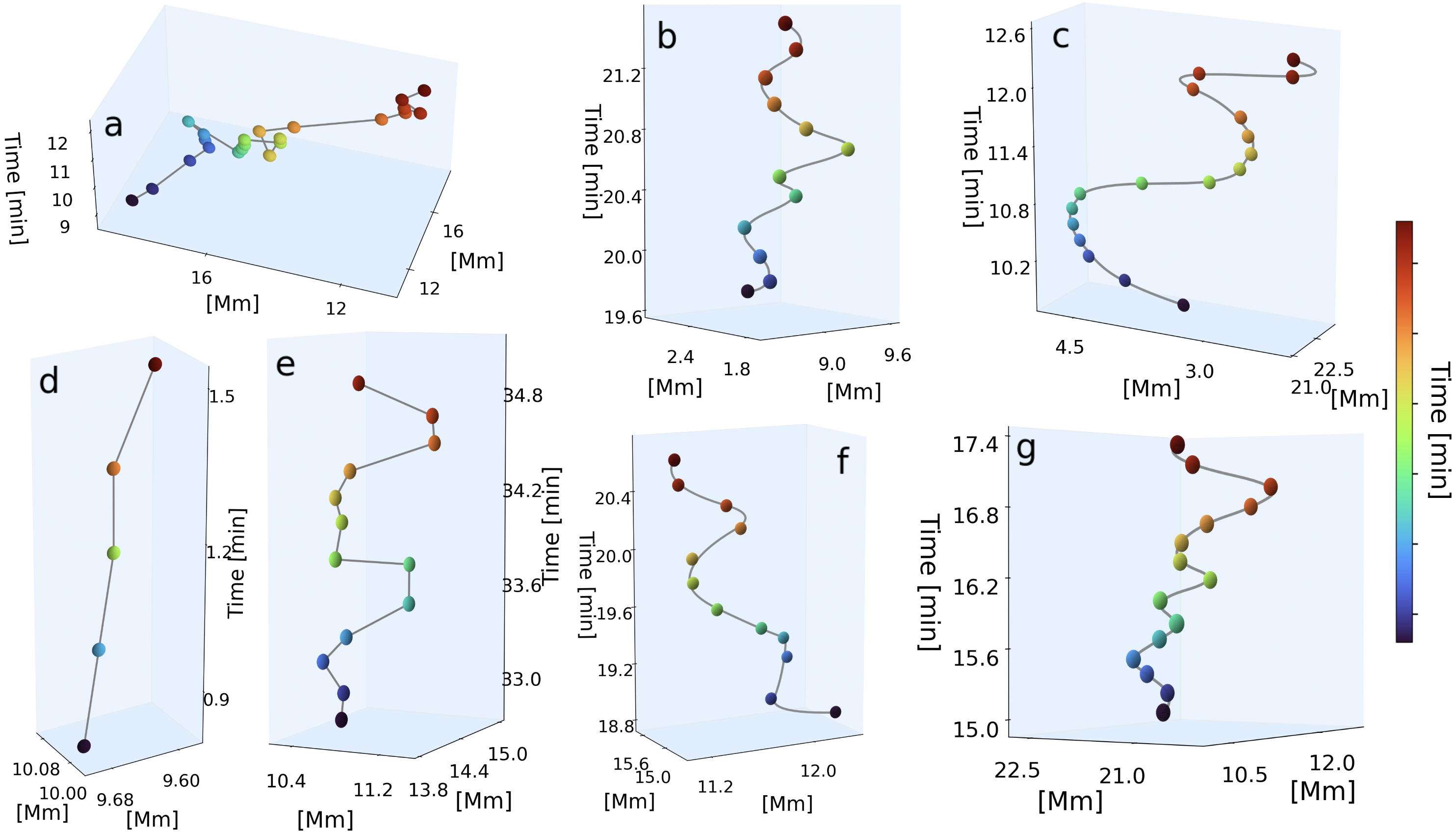}
  \end{adjustbox}
  \caption{Community centroid trajectories over time in a fixed horizontal plane. Panels b, c, f and g show four examples exhibiting helical motion, whilst panels d and e show examples of directed and kinked motion, respectively. Finally, panel a shows an unclassified centroid trajectory, exhibiting no clear motion pattern. Spheres denote centroid positions and are coloured by time.}
\label{fig:Cent_Traj}
\end{figure*}

We infer community motion from the centroid time series by computing the displacement of the centre between successive 10s frames. We classify the resulting trajectories into four types: helical, kinked, directed and no pattern. Figure \ref{fig:Cent_Traj} shows examples of each class. Helical trajectories exhibit swirling motion (b, c, f and g), kinked trajectories show abrupt changes in direction (e) and directed trajectories follow a smooth, coherent path (d). Trajectories that do not match these behaviours are grouped as no pattern and show no consistent structure (a).

We track centroid positions for coherent vortex communities across the full time interval and field of view described in Section \ref{sec:Bifrost}, evaluated at a photospheric height of $\approx 0.45$~Mm. Community motion is classified using three trajectory metrics: net displacement, total displacement and turning behaviour. We define $E$ as the ratio of net displacement to total displacement and $\bar{\theta}$ as the mean angle between successive displacement vectors. Directed trajectories are defined by $E \geq 0.9$, kinked trajectories by $\bar{\theta} > 1.5$ rad and helical trajectories by $E < 0.4$ and $\sum \theta > 4\pi$ (with $\theta$ in radians). Communities that do not satisfy any criterion are assigned to the no pattern class.

Table \ref{tab:helical} summarises the prevalence of the different centroid motion types exhibited by vortex communities. Classification requires at least four detected frames, corresponding to a 40s trajectory. Longer lived communities ($\geq 1.5$ min), which contain nine or more position vectors, provide a clearer measure of global motion and support more robust categorisation. Because these longer-lived communities persist for longer, they may also be more dynamically relevant. We therefore repeat the analysis using minimum length thresholds of 4 - 8 frames and report the resulting motion type proportions after excluding communities shorter than each threshold.

Directed vortex groups are rare in our dataset, with proportions between 0\% and 9.3\%. Persistently unidirectional centroid motion is mostly confined to short-lived communities, and straight-line trajectories are uncommon overall. Kinked vortex groups occur more frequently (14.6\% to 25.3\%) but remain a minority class. Kinked motion is most prevalent when short-lived communities are included, but it is still observed when restricting the sample to longer-lived communities.

We next consider centroid trajectories that may reflect a net vortical motion. For point vortices, groups of idealised vortices can generate helical motion in the induced flow field \citep{Saffman_1992}. Our results are consistent with similar behaviour occurring for solar vortex communities. Table \ref{tab:helical} shows helical classification rates of 32\% to 58.6\%, with higher rates among longer-lived communities. The no pattern class remains substantial (26.8\% to 33.3\%), which could arise from strict classification thresholds and or from communities with weaker intra-community interactions that do not produce a clear centroid motion signature.

\begin{table*}[t!]
\centering
\begin{tabular}{lcccc}
\toprule
Minimum Community Lifetime & Helical & No Pattern & Kinked & Directed \\
\midrule
4 Frames & 32\%   & 33.3\% & 25.3\% & 9.3\% \\
5 Frames & 36.9\% & 34.6\% & 21.5\% & 6.9\% \\
6 Frames & 44.4\% & 32.4\% & 21.3\% & 1.9\% \\
7 Frames & 47.5\% & 31.7\% & 19.8\% & 1.0\% \\
8 Frames & 51.6\% & 28.0\% & 19.4\% & 1.1\% \\
9 Frames & 58.6\% & 26.8\% & 14.6\% & 0\% \\
\bottomrule
\end{tabular}
\caption{The types of trajectories exhibited by community centroids. For each row, we only consider communities that persist for at least $x$ frames.}
\label{tab:helical}
\end{table*}

\section{Discussion and Conclusion}
In this letter, we use a community detection algorithm to analyse photospheric vortices as an interacting system rather than isolated features. We analysed 37 minutes of Bifrost simulation and tracked vortices through their spatial and temporal extent. We use the reduced network, constructed from interaction strengths defined via the Biot-Savart induced velocity, to identify key structures within the system \citep{Saffman_1992,Nair_Taira_2015,Gopalakrishnan_Meena_2018}. We identified 340 distinct communities, with an average size of 3 spanning an average of 1.12 minutes. 



Vortices are categorised according to their role within the interaction network as peripheral, connector, or hub nodes \citep{Gopalakrishnan_Meena_2021}.  These roles describe dynamical influence: peripheral vortices act mainly within their community, connectors link communities, and hubs have the greatest total outgoing induced velocity strength. Vortex tube lengths and lifetimes of fully observed vortices classified as peripheral, connector and hub are larger than those of the full vortex population, which is consistent with these vortices persisting to greater heights and remaining coherent for longer \citep{Kitiashvili_2012, Silva_2020}. Vortices with longer lifetimes and vertical extents are expected to have a larger cumulative influence on photospheric and chromospheric dynamics \citep{Silva_2024b}.


The network-based framework reveals coherent, large-scale structures that are not apparent from individual vortex detections alone \citep{Gopalakrishnan_Meena_2018, Gopalakrishnan_Meena_2021}. The centroid trajectories of these communities capture collective dynamics that extend beyond the motion of their constituent vortices, with a substantial fraction exhibiting signatures consistent with net vortical drift, exemplified in Fig.~\ref{fig:Cent_Traj}. While analogous helical motion can be demonstrated for idealised point-vortex systems, establishing such behaviour rigorously in turbulent magnetohydrodynamic regimes remains challenging \citep{Saffman_1992, Fukumoto_2005}. Nevertheless, 32\%–58.6\% of the detected communities are classified as helical, supporting the interpretation that vortex collectives may sustain global rotational dynamics and thereby enhance their capacity to excite waves \citep{Fedun_2011,Tziotziou_2023}, moving beyond localised wave drivers.




\begin{acknowledgments}
 L.M. is grateful for the STFC studentship. V.F., G.V., I.B. and S.S.A.S. are grateful to The Royal Society, International Exchanges Scheme, collaboration with Greece (IES/R1/221095), India (IES/R1/211123) and Taiwan (IEC/R3/233017). V.F., G.V. and S.S.A.S. are grateful to the Science and Technology Facilities Council (STFC) grants ST/V000977/1, ST/Y001532/1, UKRI1165. S.S.A.S. and V.F. would like to thank the International Space Science Institute (ISSI) in Bern, Switzerland, for the hospitality provided to the members of the team on `The Nature and Physics of Vortex Flows in Solar Plasmas' and ‘Tracking Plasma Flows in the Sun’s Photosphere and Chromosphere: A Review \& Community Guide’.  
\end{acknowledgments}

\appendix
\section{Vortex Detection}\label{sec:vortex_detection}
\subsection{The Gamma Method}
The $\Gamma$ method of \citet{Graftieaux_2001} provides an automated approach to identify vortex centres and boundaries from particle image velocimetry (PIV) velocity fields. This is valuable for astrophysical datasets because it scales well to large simulation and observational volumes and can be applied across a broad range of spatial scales. The method is well established in the wider fluids literature, including terrestrial and space-based applications and is commonly used in turbulent regimes where distinguishing shear from rotation is important. Figure \ref{fig:Ind_Vel} (tiles a-g) shows examples of $\Gamma$ detected vortices. The streamlines indicate that the detected boundaries and centres coincide with locally rotating flow.

The method relies on two complementary functions: $\Gamma_1$, which identifies the vortex centre and $\Gamma_2$, which detects the vortex boundary. For a given point $P$, these functions are defined in the discrete framework as follows:

\begin{equation}
     \Gamma_1 (P) = \frac{1}{N} \sum_{S} \frac{[\boldsymbol{PM} \times \boldsymbol{U}_M] \cdot \boldsymbol{z}}{\Vert \boldsymbol{PM} \Vert \cdot \Vert \boldsymbol{U}_M \Vert} = \frac{1}{N} \sum_{S}  \sin(\theta_M),
 \end{equation}

 \begin{equation}
     \Gamma_2 (P) = \frac{1}{N} \sum_{S} \frac{[\boldsymbol{PM} \times (\boldsymbol{U}_M - \tilde{\boldsymbol{U}}_P)] \cdot \boldsymbol{z}}{\Vert \boldsymbol{PM} \Vert \cdot \Vert \boldsymbol{U}_M - \tilde{\boldsymbol{U}}_P \Vert}.
 \end{equation}

Here, the functions are evaluated over a local kernel $S$ containing $N$ discrete points, where $M$ denotes a point in $S$. The velocity at point $M$ is denoted by $\boldsymbol{U}_M$, while $\tilde{\boldsymbol{U}}_P$ is the local convection velocity around point $P$, defined as:

\begin{equation}
     \tilde{\boldsymbol{U}}_P = \frac{1}{N} \sum_{M \in S} \boldsymbol{U}_M.
\end{equation}

In this context, $ \boldsymbol{z} $ is the unit vector normal to the measurement plane, $\boldsymbol{PM}$ is the vector from point $P$ to point $M$, and $\theta_M$ is the angle between $\boldsymbol{PM}$ and $\boldsymbol{U}_M$.

The functions $\Gamma_1$ and $\Gamma_2$ are based on local flow topology. This makes them less sensitive to absolute velocity magnitude and helps suppress small-scale turbulent fluctuations. Using user-defined thresholds, the method identifies candidate vortex centres and boundaries. A detection is recorded only when a $\Gamma_1$ centre lies within a $\Gamma_2$ boundary and when the absolute values exceed the chosen thresholds $\Gamma_{1_{\text{Thr}}}$ and $\Gamma_{2_{\text{Thr}}}$:
$$
     |\Gamma_1| > \Gamma_{1_{\text{Thr}}} \quad \text{and} \quad |\Gamma_2| > \Gamma_{2_{\text{Thr}}}.
$$
Large magnitudes of $\Gamma_1$ and $\Gamma_2$ (both bounded in $[-1,1]$) are associated with rotational flow, while the sign indicates the orientation of rotation.

Here, we use adaptive thresholding to better capture vortices during decay, when their rotational signature can weaken. In these phases, fixed threshold methods can miss detections and truncate lifetimes. Our dual aspect approach is designed to reduce this effect and support more consistent tracking through a vortex lifetime.

\subsection{The dual-part Gamma Method}

The dual part $\Gamma$ method builds on the original $\Gamma$ framework and uses the same $\Gamma_1$ and $\Gamma_2$ functions, but changes how they are applied during tracking. The key idea is to vary threshold values dynamically when following vortex evolution in time and height. In regions that were previously flagged as detections, we lower the $\Gamma_1$ threshold to support continued identification of the same vortex through weaker phases of its lifetime and vertical extent. In regions with no prior detection, we keep $\Gamma_1$ higher to reduce false positives. We use an initial threshold of 0.75, and when a previous detection exists, we reduce it to $2/\pi$.

For boundary identification, we fix $\Gamma_{2_{\text{Thr}}} = 2/\pi$ for both initial and subsequent detections. This follows \citet{Graftieaux_2001}, who showed that in the limit $S \rightarrow 0$, regions with $|\Gamma_2| > 2/\pi$ are rotation dominated rather than dominated by shear or strain. Keeping the boundary threshold constant helps keep the detected vortex area comparable over time, aside from physical changes during vortex evolution. Given that our calculations for circulation and thus induced velocity are based upon the boundary of the vortex, the constant boundary threshold is particularly important in this work. It ensures consistency in circulation estimates across the considered vortices, creating a more reliable progression of vortex community detections. 

\subsection{Normalisation as a Preprocessing Step}

Robust vortex detection is central to our analysis. To improve the boundary estimates, we apply a normalisation pre-processing step to the horizontal velocity field and compute the $\Gamma$ fields from this normalised field. This removes sensitivity to velocity magnitude and emphasises flow topology, which can lead to more consistent vortex boundaries. In particular, this step produces more convex boundaries \citep{McClure_2026}, which have been used as an indicator of more reliable vortex identification \citep{Haller_2016, Tziotziou_2023}.

\section{Peripheral, Connector and Hub Definitions}
In this paper, we use the peripheral, connector and hub definitions to classify vortices (nodes) within the interaction network \citep{Gopalakrishnan_Meena_2021}. The characteristic properties of these structures are described in Section \ref{sec:vortex_interactions}. These classes are defined relative to the reduced network structure and, therefore, relative to the inferred vortex interaction strengths.

Hub nodes are characterised by a large total outward inducing strength \citep{Gopalakrishnan_Meena_2021}. For vortex $i$, we define
\begin{equation}
s_i = \sum_{j \neq i} u_{i \rightarrow j}.
\end{equation}
In our analysis, we identify a single hub node, defined as the vortex with the maximum outward inducing strength, $i_{\rm hub} = \arg\max_i s_i$.

Peripheral and connector nodes are identified using community structure by comparing how strongly a vortex interacts within its assigned community versus outside it. For a vortex $i$ in community $k$, we define the intra-community interaction strength as
\begin{equation}
s_i^{\rm intra} = \sum_{\substack{j \in C_k \ j \neq i}} u_{i \rightarrow j},
\end{equation}
and the inter-community interaction strength as
\begin{equation}
s_i^{\rm inter} = \sum_{j \notin C_k} u_{i \rightarrow j}
\end{equation}
\citep{Gopalakrishnan_Meena_2021}. These quantities quantify the relative influence of a vortex on its own community compared to the rest of the network. The balance between $s_i^{\rm intra}$ and $s_i^{\rm inter}$ varies between vortices depending on their circulation and spatial distribution.

We introduce a within-module $Z$ score to quantify the relative intra-community interaction strength of a vortex within its assigned group \citep{Gopalakrishnan_Meena_2021}. For a vortex $i$, this is defined as
\begin{equation}
Z_i = \frac{s_i^{\rm intra} - \overline{s^{\rm intra}}_{k}}{\sigma_{s^{\rm intra}_k}}.
\end{equation}
Here, $\overline{s^{\rm intra}}_{k}$ and $\sigma_{s^{\rm intra}_k}$ are the mean and standard deviation of $s^{\rm intra}$ computed over all vortices in the community $k$ to which vortex $i$ belongs.

Finally, we introduce a participation coefficient, $P$, to quantify how a vortex’s inter-community influence is distributed across the network \citep{Gopalakrishnan_Meena_2021}. This measures how evenly a vortex spreads its interaction strength across communities. With $P \in [0,1]$, its high values indicate that a vortex’s influence is distributed relatively evenly across communities, whereas the low values indicate that its influence is concentrated mainly within its assigned community. For a vortex $i$ in community $k$,
\begin{equation}
P_i = 1 - \left[ \left( \frac{s_i^{\rm intra}}{s_i}\right)^2 + \sum_{j \notin C_k} \left( \frac{u_{i\rightarrow j}}{s_i} \right)^2 \right].
\end{equation}

For each community $k$, both the peripheral and connector vortices are selected as the vortex with the largest within module $Z$ score, i.e. $\max_{i,, i \in C_k} Z_i$. The distinction between peripheral and connector then depends on how strongly the community interacts with the rest of the network. We define the peripheral vortex as the $Z$-maximising vortex in the community with the smallest mean participation coefficient, $\min_k \overline{P_k}$ and the connector vortex as the $Z$-maximising vortex in the community with the largest mean participation coefficient, $\max_k \overline{P_k}$ \citep{Gopalakrishnan_Meena_2021}. Here, $\overline{P_k}$ is the mean value of $P$ across all vortices in the community $k$.

In the solar context, where large domains can be sparsely populated, the optimal network partition can include single vortex communities. For these communities $s^{\rm intra}=0$, $s^{\rm inter}$ is typically small and have undefined $Z$. We treat these cases as isolated communities and exclude them from the peripheral, connector and hub classification.

\bibliography{mybibliography}{}
\bibliographystyle{aasjournal}

\end{document}